\documentclass[twocolumn,showpacs,preprintnumbers,amsmath,amssymb,prl]{revtex4}

\usepackage{graphicx}
\usepackage{amssymb}

\begin{document}


\title{Viscosity anomaly in core-softened liquids}



\author{Yu. D. Fomin}
\affiliation{Institute for High Pressure Physics, Russian Academy
of Sciences, Troitsk 142190, Moscow, Russia}

\author{V. N. Ryzhov}
\affiliation{Institute for High Pressure Physics, Russian Academy
of Sciences, Troitsk 142190, Moscow, Russia}
\affiliation{Moscow
Institute of Physics and Technology, 141700 Moscow, Russia}

\date{\today}

\begin{abstract}
The present article presents a molecular dynamics study of several
anomalies of core-softened systems. It is well known that many
core-softened liquids demonstrate diffusion anomaly. Usual
intuition relates the diffusion coefficient to shear viscosity via
Stockes-Einstein relation. However, it can break down at low
temperature.  In this respect it is important to see if viscosity
also demonstrates anomalous behavior.
\end{abstract}





\pacs{61.20.Gy, 61.20.Ne, 64.60.Kw} \maketitle

\section{Introduction}

It is well known that some liquids (for example, water, silica,
silicon, carbon, and phosphorus) show an anomalous behavior \cite{
book,book1,deben2001,netz,denstr,errington2}:their phase diagrams
have regions where a thermal expansion coefficient is negative
(density anomaly), self-diffusivity increases upon compression
(diffusion anomaly), and the structural order of the system
decreases with increasing pressure (structural anomaly)
\cite{deben2001,netz}. A number of studies demonstrates water-like
anomalies in fluids that interact through spherically symmetric
potentials (see, for example, \cite{buld2009,wepre,wepre1} and
references therein). Many of these studies report the appearance
of diffusion anomaly in different systems. However, the diffusion
coefficient is closely related to shear viscosity of liquid
therefore one can expect that the shear viscosity also
demonstrates some kind of anomalous behavior.

Although many studies of core-softened systems report the
diffusivity calculations there is a lack of studies which
calculate shear viscosity. This can be related to the fact that
viscosity is much harder to compute in simulation. So the usual
intuition is applied: the viscosity can be extracted from the
diffusion coefficient by Stockes-Einstein (SE) relation
\cite{hansen}. However, it was recently found that SE relation can
be violated at low temperatures \cite{sev1,sev2}. This case the
usual intuition can fail to predict the viscosity behavior
correctly. In this respect it is important to monitor both the
diffusion coefficient and shear viscosity of core-softened liquids
at low temperatures to see theirs behavior in the regions of
anomalous behavior.

The goal of the present article is to investigate the behavior of
shear viscosity of core-softened fluids at low temperatures, to
see if theirs shear viscosity demonstrates anomalous behavior and
if so to find the relation between the viscosity anomaly region
and the regions of other anomalies.

\section{Systems and methods}

Two systems are studied in the present work. The first one is a
core-softened system introduced by de Oliveira et al
\cite{barbpot}. This system is described by the Lennard-Jones
potential with Gaussian well (LJG):

\begin{equation}
  U(r)=4\varepsilon
  \left[\left(\frac{\sigma}{r}\right)^{12}-\left(\frac{\sigma}{r}\right)^{6}\right]+a\varepsilon
  \cdot \exp\left(-\frac{1}{c^2}\left(\frac{r-r_0}{\sigma_0}\right)^2\right),
\end{equation}
with $a=5.0$, $r_0/ \sigma=0.7$ and $c=1.0$. The diffusivity of
this system was studied in several papers
\cite{barbpot,indiabarb,werosbreak,wetraj,wetraj1}. Note that the
parameters of the potential are chosen in such a way that the
effect of attraction becomes negligibly small and one can consider
this system as a purely repulsive core-softened one.

The second system studied in this work is Soft Repulsive Shoulder
System (SRSS) introduced in the work \cite{wejcp}. The potential
of this system has form:

\begin{equation}
  U(r)=\varepsilon
  \left(\frac{\sigma}{r}\right)^{14}+\frac{1}{2}\varepsilon
  \cdot[1-\tanh(k_0\{r-\sigma_1\})],
\end{equation}
where $\sigma$ is "hard"-core diameter, $\sigma_1=1.35$ is
soft-core diameter and $k_0=10.0$. In Ref. \cite{wepre} it was
shown that this system demonstrates anomalous behavior. Our later
publications gave detailed study of diffusion, density and
structural anomalies in this system \cite{wepre1}.

It is well known that there is a close link between diffusion
coefficient and shear viscosity of liquids. Viscosity is a
quantity which is usually measured in experiments. However, due to
the technical problems the simulation of shear viscosity
represented in the literature is very poor. One of the goals of
this article is to study the behavior of viscosity of the two
model liquids described above. Taking into account that the
diffusion coefficient demonstrates anomalous behavior for these
systems we are interesting to see if the viscosity also
demonstrates some anomalies.

In order to study the transport coefficients of the systems we
used Molecular Dynamics method. In both cases a system of $N=1000$
particles was simulated. The equations of motion were integrated
by velocity Verlet algorithm. In case of LJG system the time step
was set to $dt=0.001$, the equilibration period was $3.5 \cdot
10^6$ steps and the production period $1 \cdot 10^6$ steps. In the
case of SRSS the equilibration the time step was $dt=0.0005$, the
equilibration period was $3.5 \cdot 10^6$ and the production run
was $0.5 \cdot 10^6$ steps. The cut-off radius was set $3.5$ for
LJG system and $2.2$ for SRSS. Velocity rescaling was applied
during equilibration, the production corresponded to $NVE$
ensemble. Shear viscosity is difficult to measure in simulation
because of large fluctuations of shear stress function. In order
to improve the precision of the data we increased the
equilibration time in anomalous region up to $7.5 \cdot 10^6$ and
the production time up to $1.5 \cdot 10^6$ for some simulations.

In order to get good statistics on the transport properties of the
systems many data points were simulated. In the case of LJG system
the data points were chosen in the density interval from
$\rho=0.05$ till $\rho=0.3$ with step $\delta \rho =0.01$ along
several isotherms. The following isotherms were considered:
$T=0.15;0.2;0.25;0.3;0.4;0.5;1.0$. In order to see the anomalous
region better we also simulated the isotherms $T=0.17$ and $0.23$
for the densities from $\rho=0.08$ up to $0.18$ with step $0.01$.

In case of SRSS we used the densities from $\rho=0.3$ up to
$\rho=0.8$ with step $\delta \rho =0.05$ and temperatures
$T=0.2;0.25;0.3;0.35;0.4;0.5;0.7$ and $1.0$.

The diffusion coefficients were computed via Einstein relation and
shear viscosity by integration of shear stress autocorrelation
function.

\section{Results and discussion}

\subsection{Lennard-Jones - Gauss system}

As it was mentioned in the introduction the viscosity anomaly in
LJG system was already reported in Ref. \cite{egorov}. Here we
make a more detailed simulation study of this anomaly. Our goal is
to see the location of the anomaly in $\rho-T$ plane and its
relation with other anomalies, such as diffusion anomaly, density
anomaly and structural anomaly.

\begin{figure}
\includegraphics[width=8cm, height=8cm]{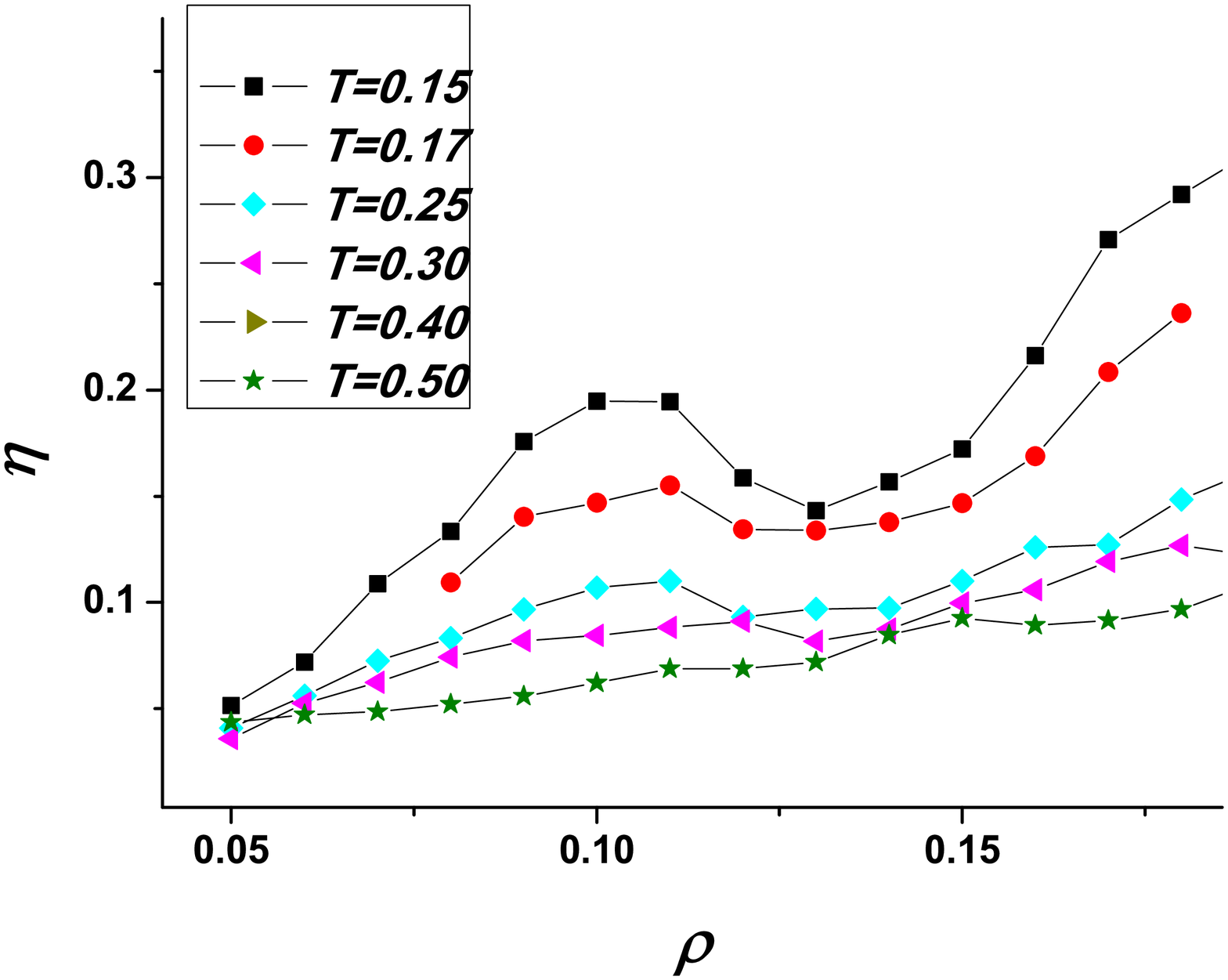}%

\caption{\label{fig:fig1} (Color online). Viscosity anomaly in JLG
system.}
\end{figure}

Figs.~\ref{fig:fig1} shows the viscosity along several isotherms
for LJG system. One can see that the anomaly is very pronounced
for the temperature $T=0.15$, but it rapidly disappears with
increasing temperature. At $T=0.3$ the anomaly is already of the
order of numerical accuracy and we estimate this temperature as
the temperature where viscosity anomaly disappears.

\begin{figure}
\includegraphics[width=8cm, height=8cm]{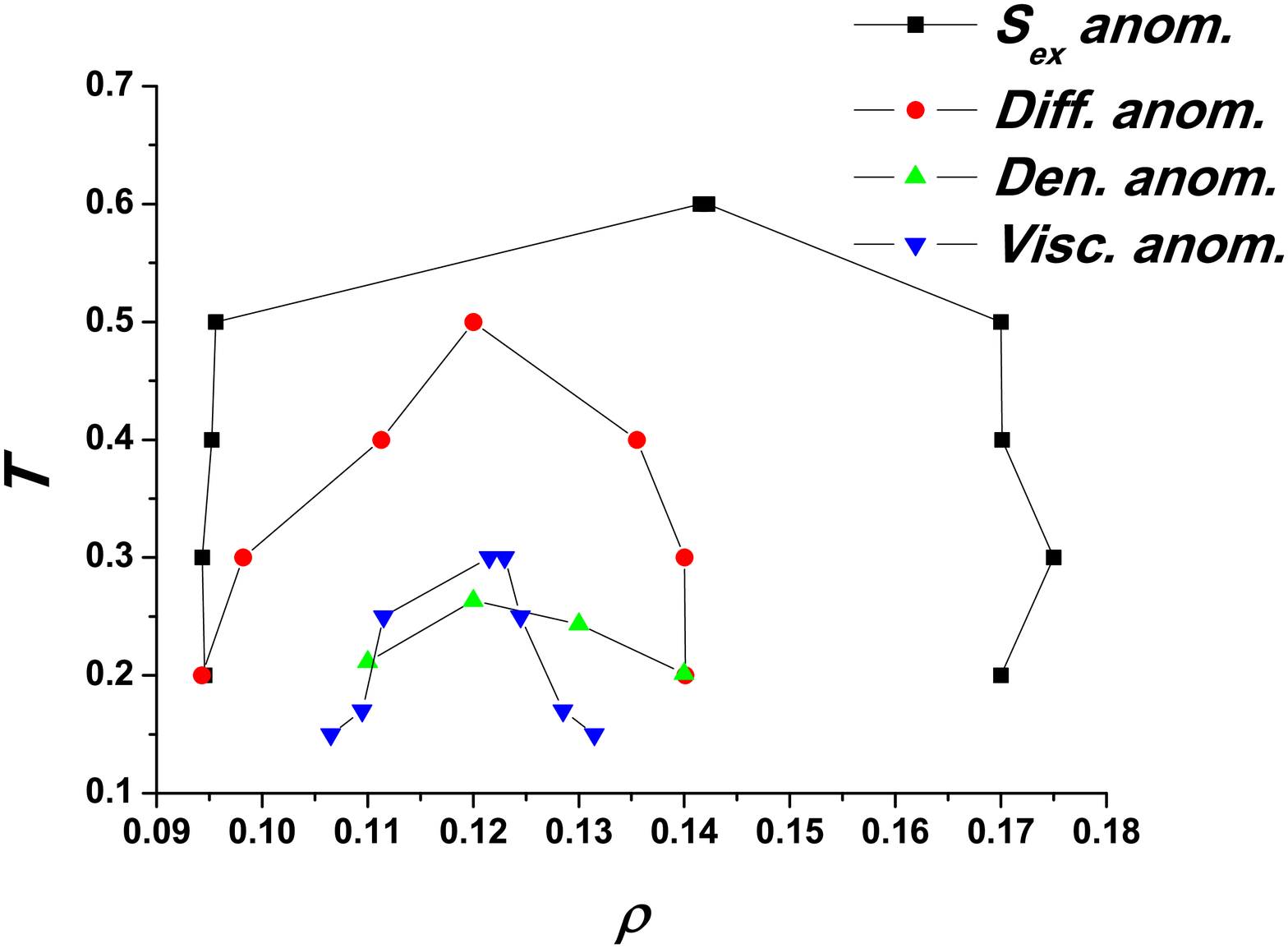}%

\caption{\label{fig:fig2} (Color online). Location of anomalous
regions in $\rho-T$ plane for LJG system.}
\end{figure}

The location of diffusion, density and structural anomalies of LJG
system in $\rho-T$ plane have already been reported in literature
\cite{indiabarb}. Figs.~\ref{fig:fig2} shows the regions of these
anomalies and the viscosity anomaly. Interestingly, as it was
proposed in Ref. \cite{debenedetti} the anomalous regions are
enveloped in each other. However, the viscosity anomaly violates
this rule: it has partial overlap with density anomaly, but no one
of them is inside of one another. It was also shown in the
literature that from the thermodynamic arguing it follows that the
density anomaly region is always inside the structural anomaly
one, while the diffusion anomaly can have any location with
respect to the other anomalies \cite{denstr,denstr1}. The
viscosity anomaly is another example of anomalies of dynamic
rather then thermodynamic properties. Therefore, one can expect
that the viscosity anomaly can also have any possible location
with respect to the density and structure anomalies.

In our previous work we showed that the anomalies can be visible
along some paths in thermodynamic space while along others they
can be invisible \cite{werosbreak,wetraj,wetraj1}. For example,
diffusion anomaly is seen along isotherms but not isochors. An
important consequence of this difference is that Rosenfeld excess
entropy scaling for diffusion coefficient \cite{ros,ros1} is
fulfilled along isochors but breaks down along isotherms
\cite{wepre1}. This makes important to see the viscosity behavior
along different trajectories.

\begin{figure}
\includegraphics[width=8cm, height=8cm]{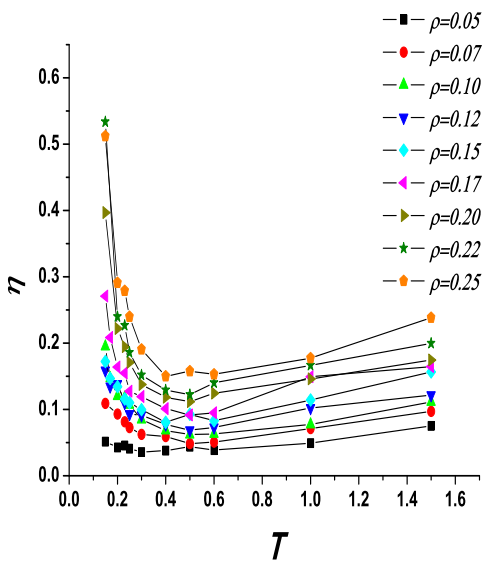}%

\caption{\label{fig:fig3} (Color online). Shear viscosity of LJG
system along several isochors.}
\end{figure}

Figs.~\ref{fig:fig3} shows shear viscosity of LJG system along
several isochors. One can see that viscosity demonstrates a
minimum. Viscosity minimum along isobars was observed
experimentally for water \cite{visc-water}. The authors called
this minimum as "viscosity anomaly". However, in our previous work
we showed that viscosity minimum along isochors appears naturally
because of the interplay of potential-potential and
kinetic-kinetic correlations even in simple liquids \cite{wesoft}.
The same results can be obtained for isobars (not shown in Ref.
\cite{wesoft}).

\begin{figure}
\includegraphics[width=8cm, height=8cm]{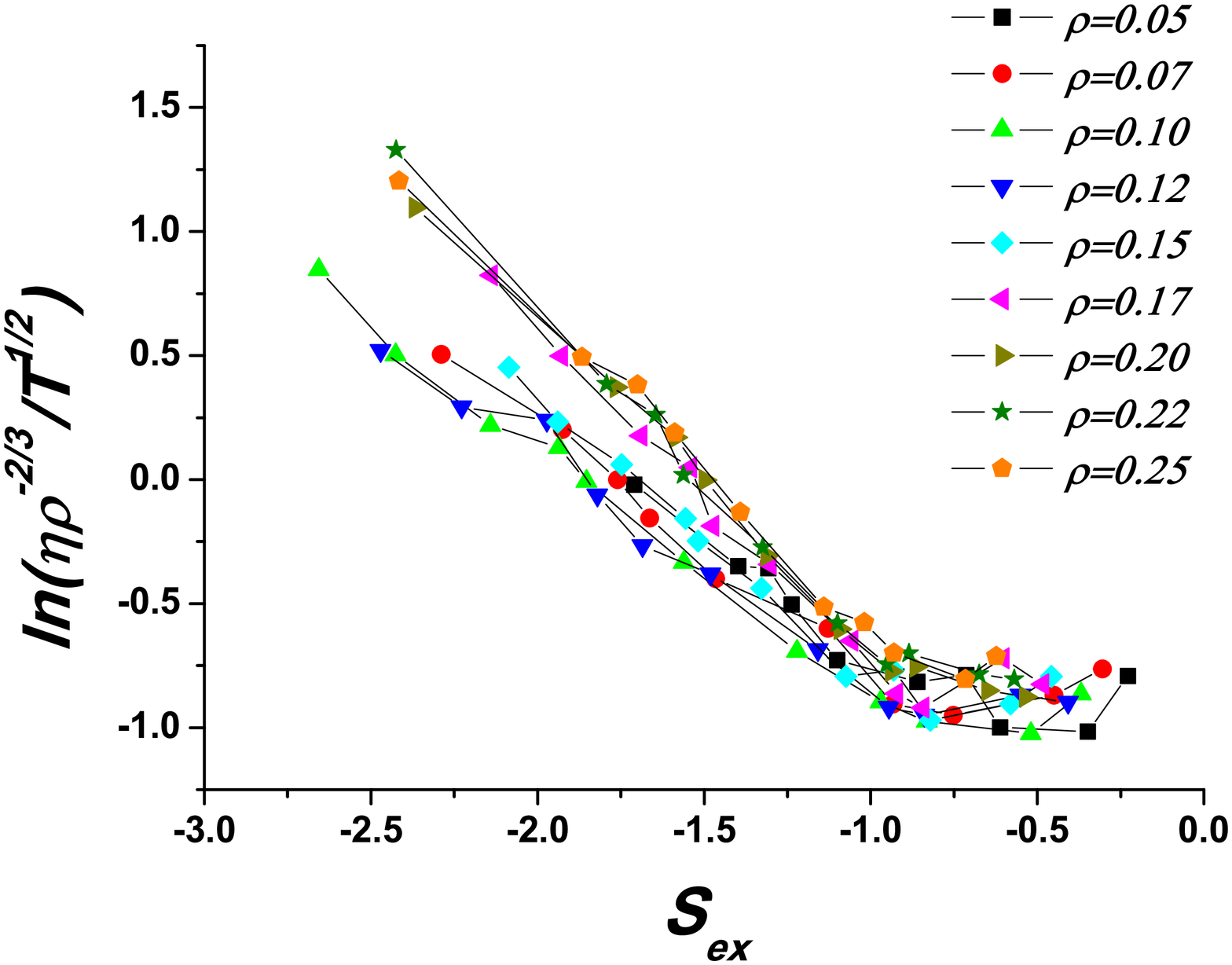}%

\caption{\label{fig:fig4} (Color online). Rosenfeld relation for
shear viscosity of LJG system along isochors.}
\end{figure}

\begin{figure}
\includegraphics[width=8cm, height=8cm]{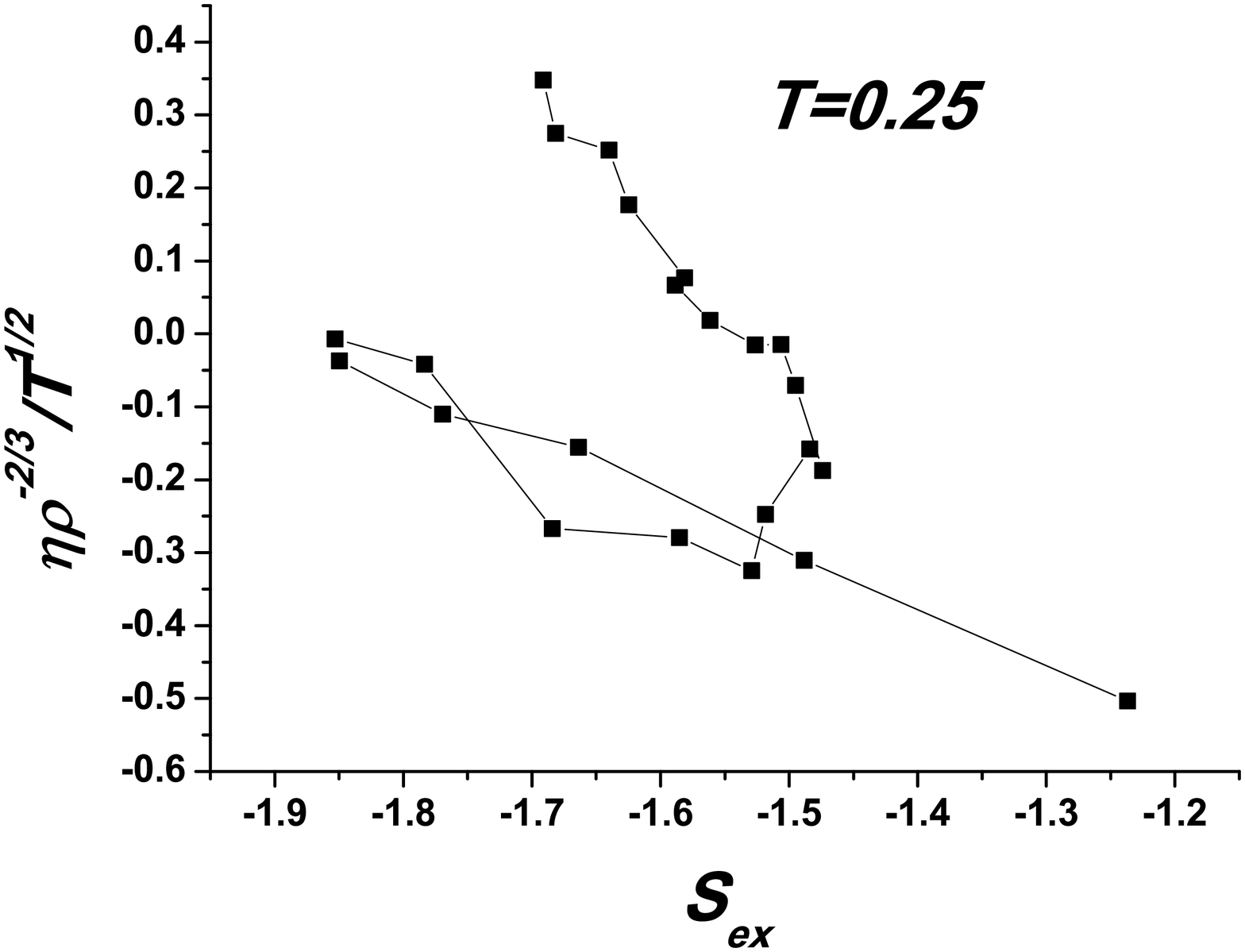}%

\includegraphics[width=8cm, height=8cm]{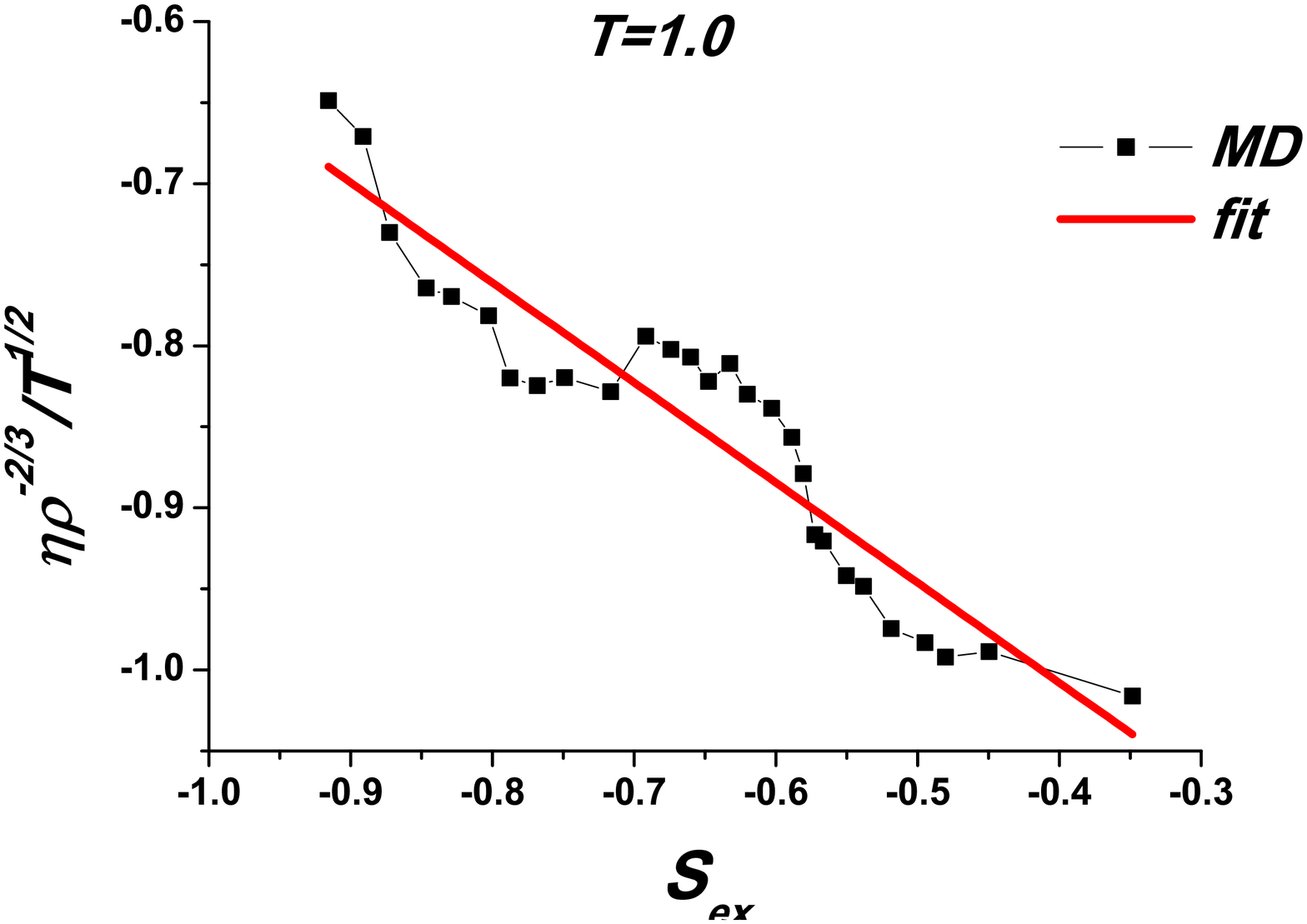}%

\caption{\label{fig:fig5} (Color online). Rosenfeld relation for
shear viscosity of LJG system along isotherms at low and high
temperature.}
\end{figure}

Figs.~\ref{fig:fig4} shows the Rosenfeld relation for shear
viscosity of LJG system along isochors. One can see that the
linear relation between $\ln(\frac{\eta \rho ^{-2/3}}{T^{1/2}})$,
which is predicted by Rosenfeld relation, holds true except the
low $S_{ex}$ region. However, if we consider the Rosenfeld
relation along isotherms we see that it breaks down at low
temperature (Figs.~\ref{fig:fig5} (a)). Here we observe a self
crossing loop like the one observed for diffusion in our previous
works \cite{wepre1,werosbreak,wetraj,wetraj1}. Fig.~\ref{fig:fig5}
(b) shows the Rosenfeld scaling for high temperature ($T=1.0$).
Points correspond to the data from simulations while the straight
line is the best fit line. One can see that the simulation points
demonstrate some kind of oscillations around the best fit line. We
relate these oscillations to the numerical inaccuracies in the
viscosity computations and we believe that Rosenfeld scaling of
viscosity does work for high enough temperatures.

One can conclude that as in the case of diffusion coefficient
Rosenfeld relation for shear viscosity in systems with
thermodynamic anomalies holds true along isochors but breaks down
along low temperature isotherms. This confirms the idea of
different behavior of a system along different trajectories in
$\rho - T - P$ space which is discussed in details in Ref.
\cite{wetraj,wetraj1}.

\subsection{Soft Repulsive Shoulder system}

The second system considered in this work is Soft Repulsive
Shoulder System (Eq. (2)). Phase diagram and anomalous behavior of
this system were studied in details in our previous works
\cite{wejcp,wepre,wepre1}. However, the shear viscosity of this
system is measured for the first time.

\begin{figure}
\includegraphics[width=8cm, height=8cm]{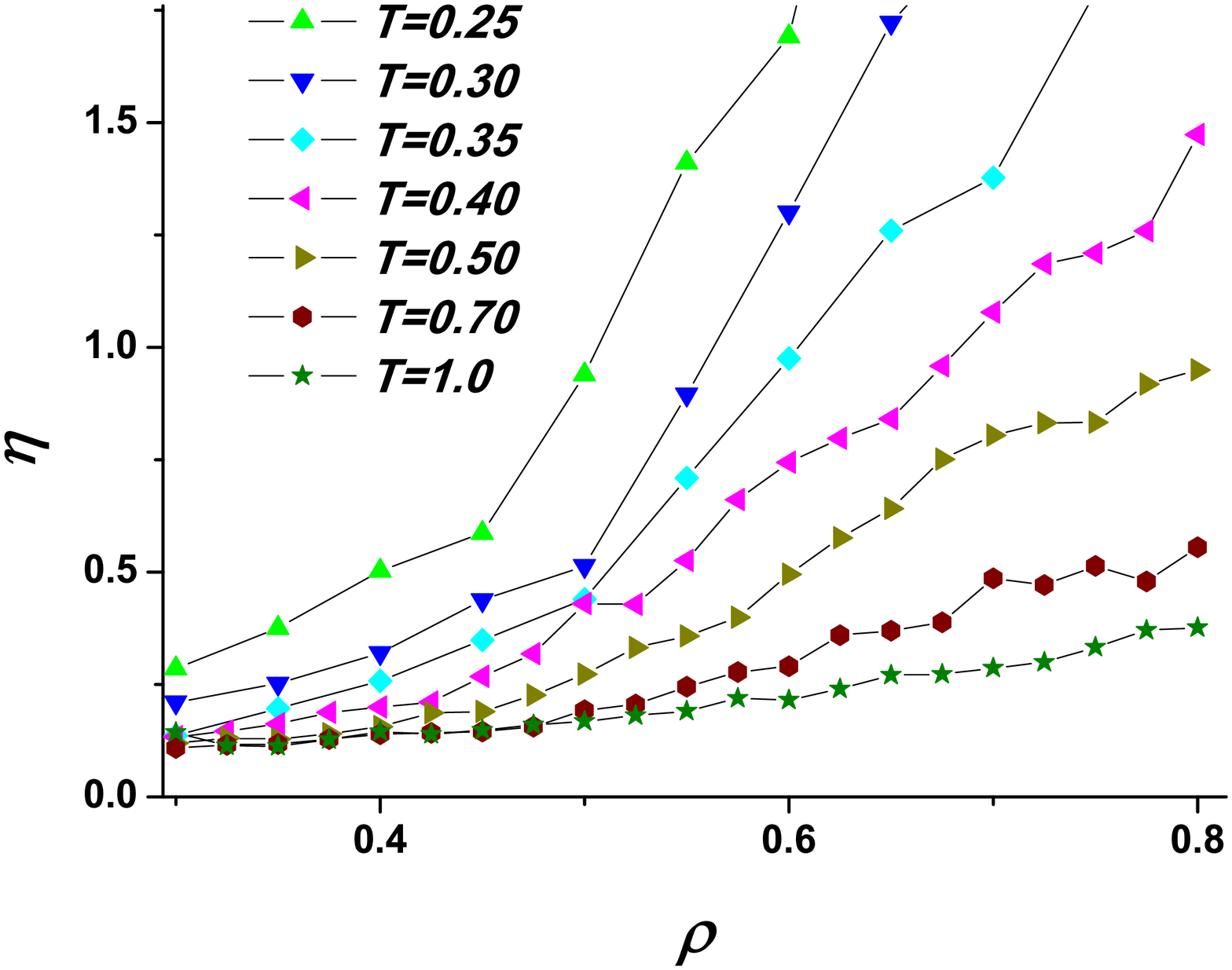}%

\includegraphics[width=8cm, height=8cm]{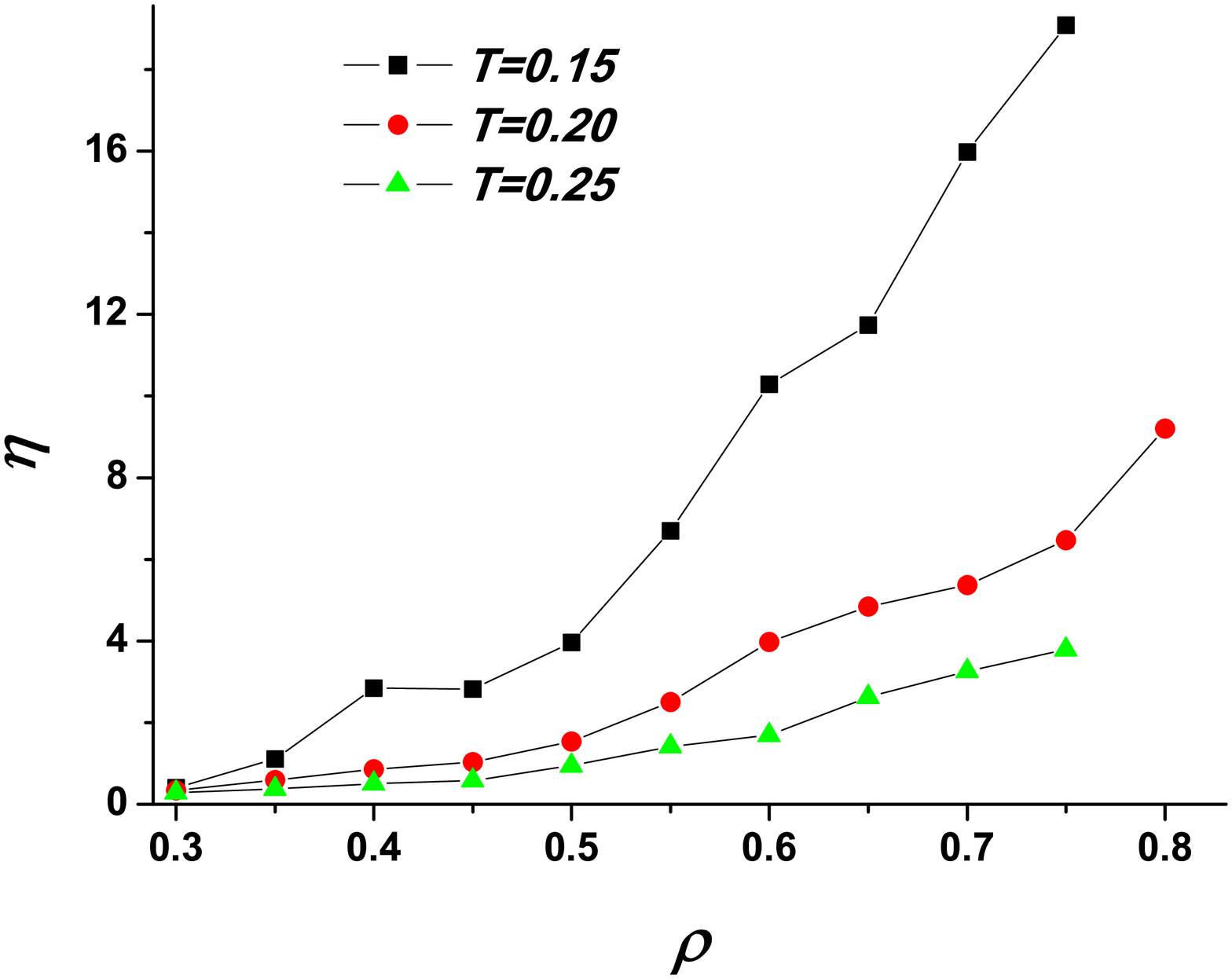}%

\caption{\label{fig:fig6} (Color online). Viscosity of SRSS system
along a set of isotherms at (a) low and (b) intermediate and high
temperatures.}
\end{figure}

Figs.~\ref{fig:fig6} (a) and (b) show the shear viscosity of SRSS
system along several isotherms. One can see that at the lowest
temperature $T=0.15$ a tiny loop develops at the densities $\rho
=0.40 - 0.45$. However, the size of this loop in inside the error
bar, so we can not consider it as a real anomaly. We believe that
the anomaly appears at lower temperatures. Though the shear stress
autocorrelation function decays very slowly and viscosity
calculations become very difficult.

\begin{figure}
\includegraphics[width=8cm, height=8cm]{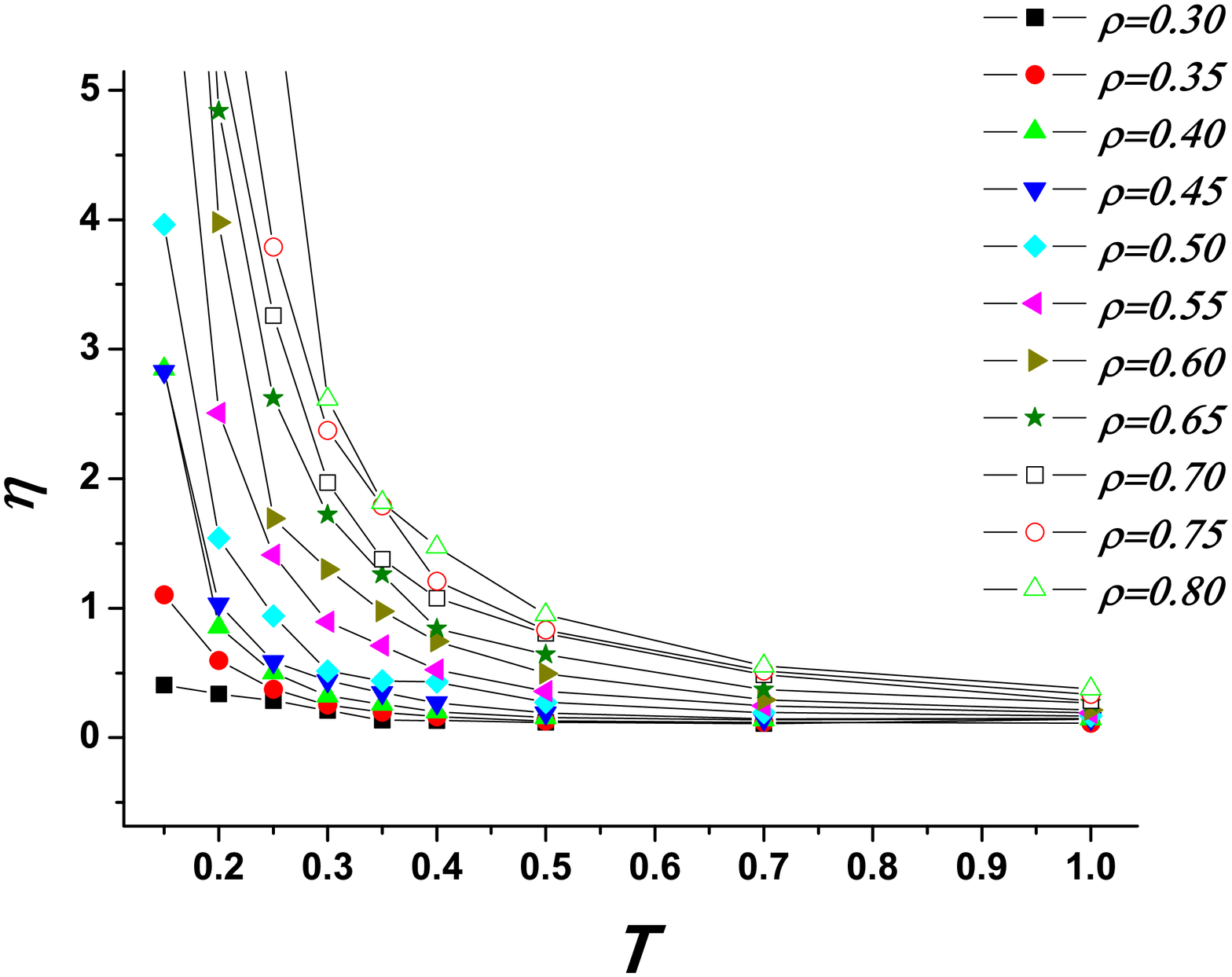}%

\caption{\label{fig:fig7} (Color online). Viscosity of SRS system
along a set of isochors.}
\end{figure}

Figs.~\ref{fig:fig7} shows the shear viscosity plotted along a set
of isochors. One can see that for all presented densities the
viscosity curves monotonically decrease with increasing
temperatures. Basing on the arguments of our previous work
\cite{wesoft} we expect that shear viscosity passes a minimum at
higher temperatures.

\begin{figure}
\includegraphics[width=8cm, height=8cm]{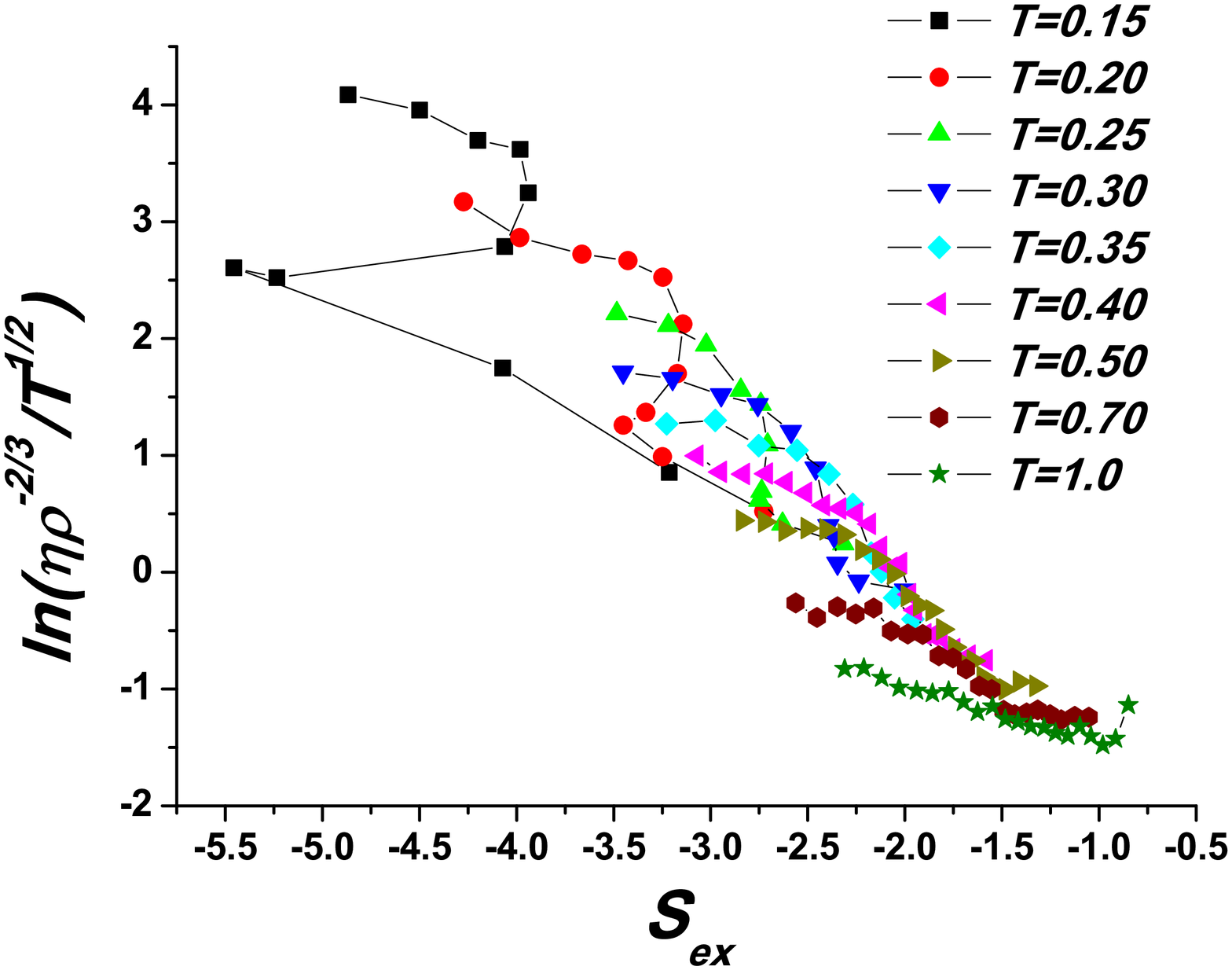}%

\includegraphics[width=8cm, height=8cm]{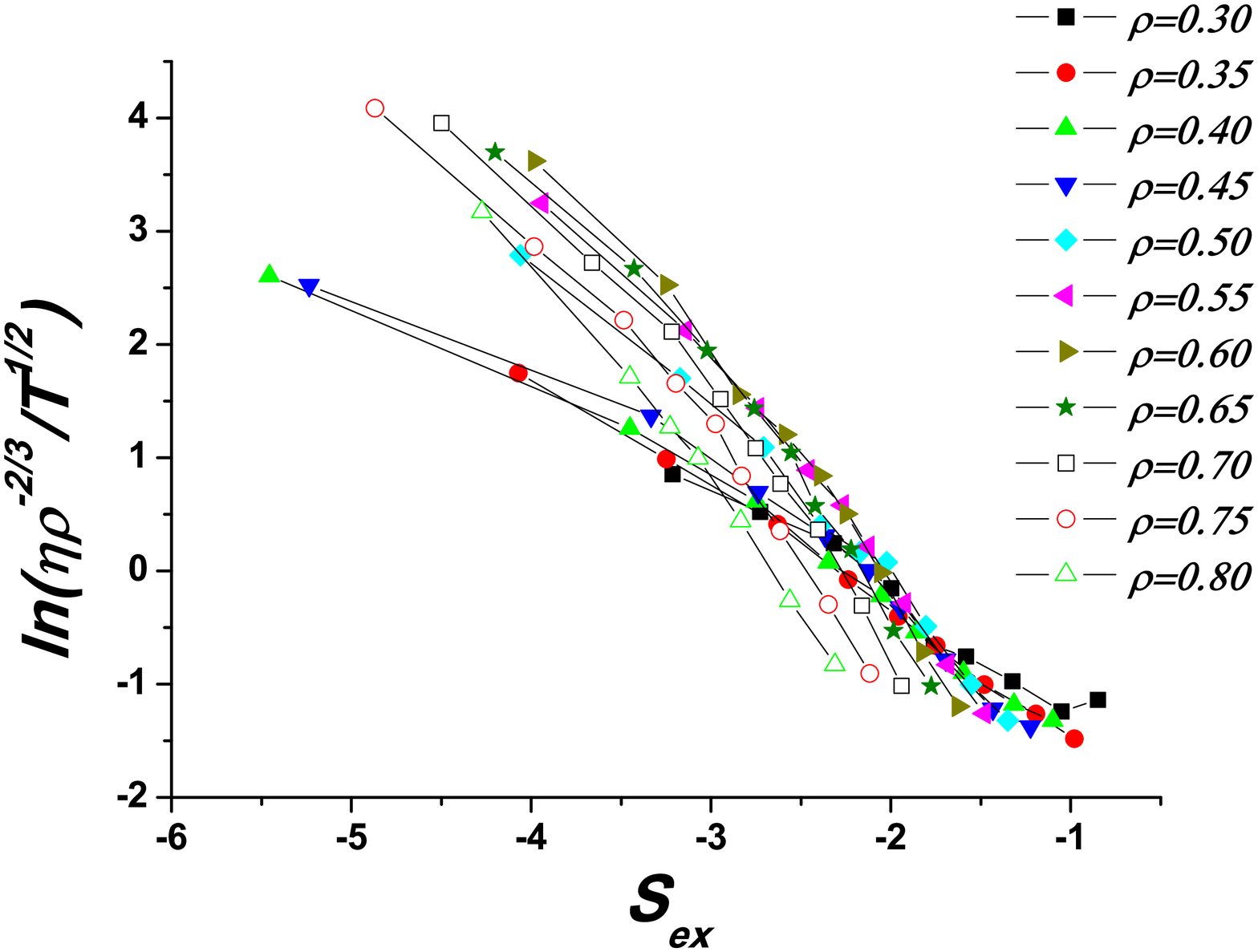}%

\caption{\label{fig:fig8} (Color online). Rosenfeld scaling plots
for SRS along (a) isotherms and (b) isochors.}
\end{figure}

Figs.~\ref{fig:fig8} (a) and (b) show the Rosenfeld scaling plots
for SRSS along isotherms and isochors. One can see that the
scaling relations break down in case of isotherms. The reason for
this breakdown is that while viscosity is monotonous function of
density the excess entropy demonstrates anomalous behavior
\cite{wetraj,wetraj1}. As a result viscosity as function of
$S_{ex}$ demonstrates nonlinear behavior.

At the same time Rosenfeld relation holds true along isochors
(Figs.~\ref{fig:fig8} (b)). All isochors can be divided in
low($\rho \leq 0.45$) and high density ($\rho > 0.45$) groups. The
curves belonging to the same groups have similar slopes while the
slope of the curves from different groups is essentially
different. The reason for this change is that at low densities the
system can be essentially approximated by the system with
effective diameter $\sigma$ while at high densities with diameter
$d$. This change of particle size alters also the kinetic and
thermodynamic properties of the system which we observe as the
slope change in Figs.~\ref{fig:fig8} (b).

\subsection{Stokes-Einstein Relation}

In our previous work \cite{wetraj,wetraj1} we showed that the
discrepancy in the diffusion and structural anomalies regions
leads to the Rosenfeld relation breakdown along isotherms. As it
was shown in the previous section the regions of diffusion and
viscosity anomalies are also different. In this respect it becomes
important to see if Stockes-Einstein relation still holds true in
the anomalous regions.

The Stockes-Einstein relation can be written in the following
form:

\begin{equation}
  c_{SE}=\frac{k_BT}{\pi D \eta d} \approx const,
\end{equation}
where $d$ is the character particle size. The coefficient $c_{SE}$
should be approximately constant and belong to the interval $2\leq
c_{SE} \leq 3$. The limiting values $c_{SE}=2$ and $c_{SE}=3$
correspond to the stick and slip boundary conditions.

\begin{figure}
\includegraphics[width=8cm, height=8cm]{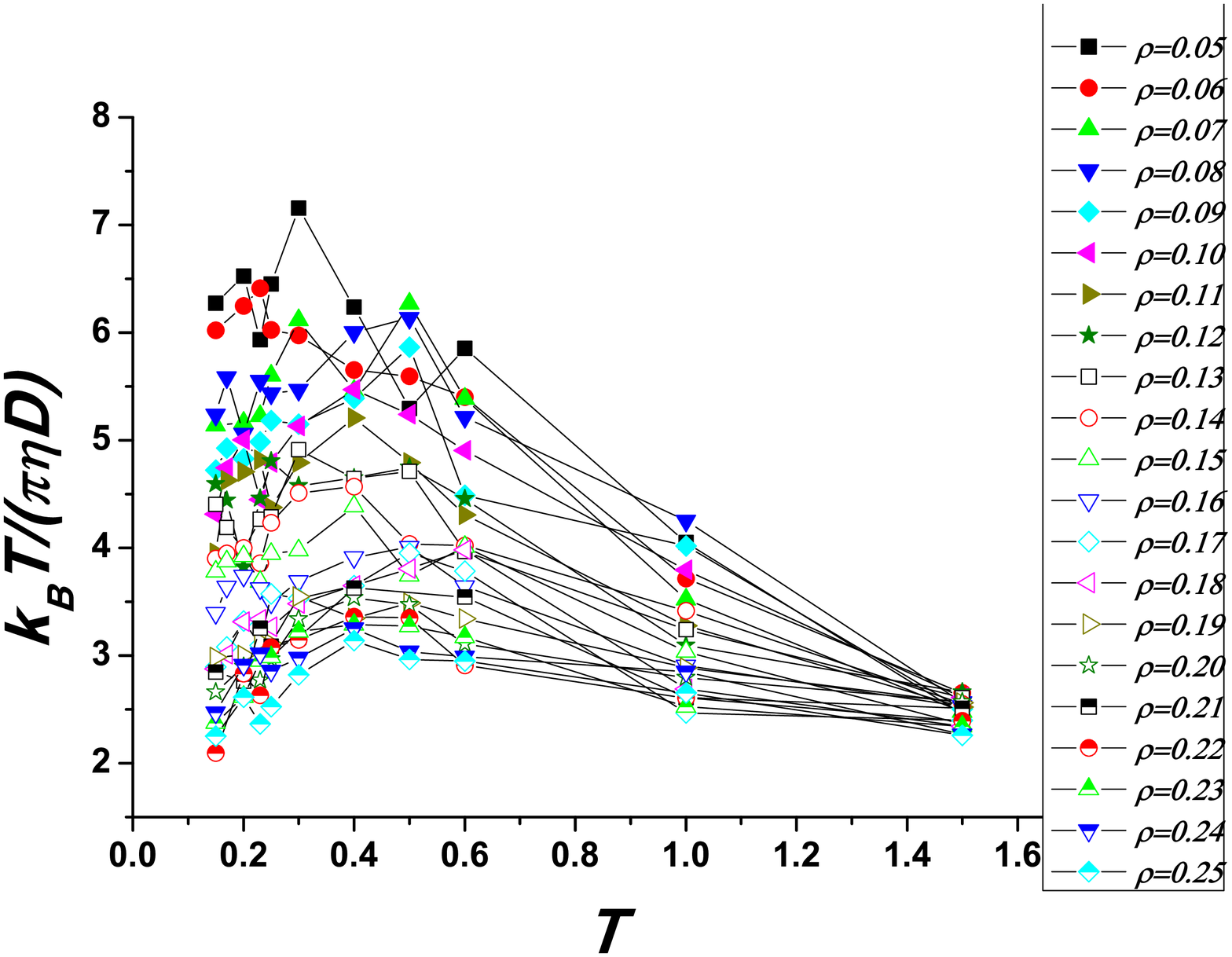}%

\includegraphics[width=8cm, height=8cm]{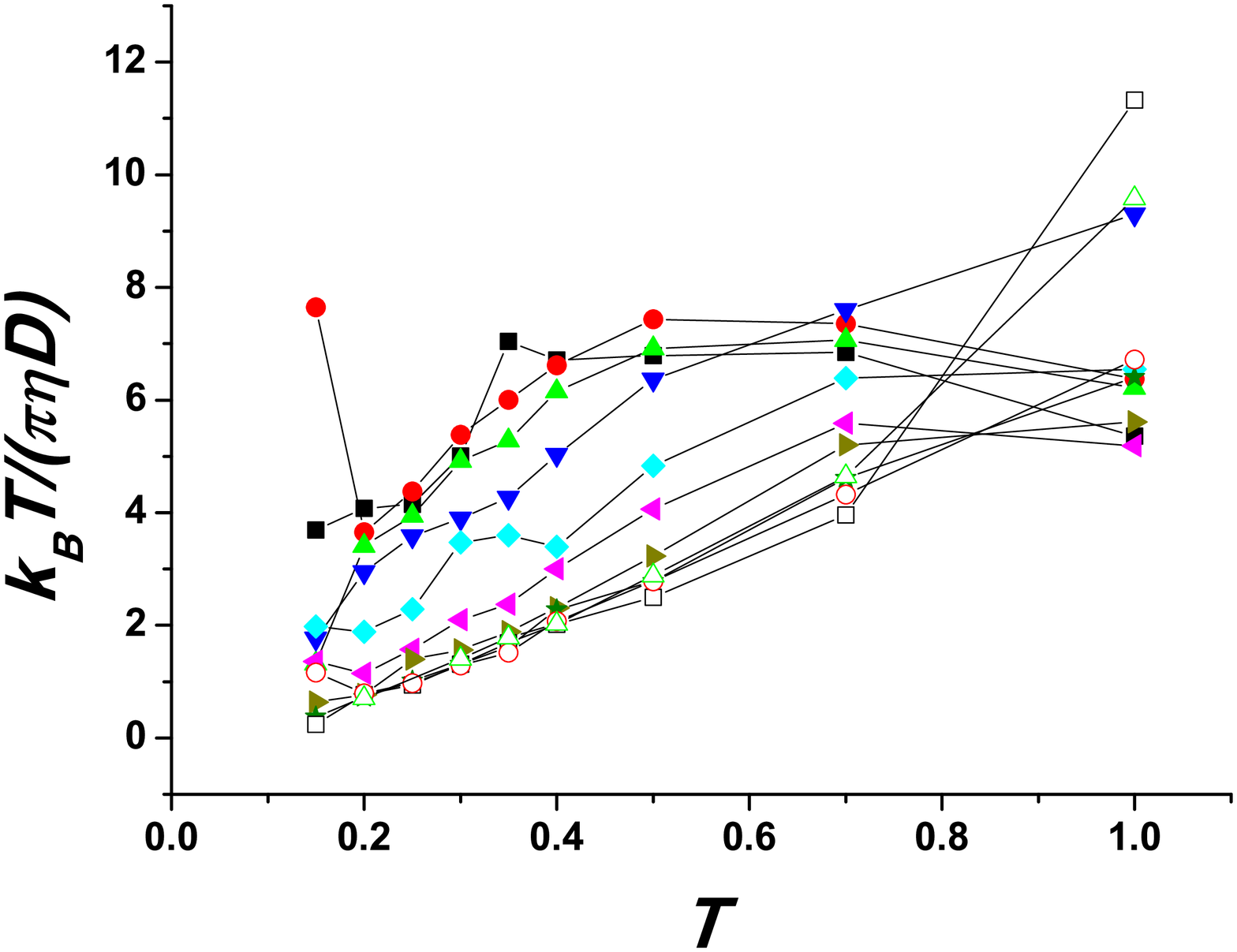}%

\caption{\label{fig:fig9} (Color online). $c_{SE}$ for (a) LJG
system and (b) SRS along several isochores.}
\end{figure}

Figs.~\ref{fig:fig9} (a) and (b) show the Stockes-Einstein
coefficient for LJG and SRSS systems along isochors. One can see
that in both cases $c_{SE}$ is not constant. At the same time the
numerical values of $c_{SE}$ in both cases can be far from the
interval $[2,3]$. In our previous work we studied Stockes-Einstein
relation for Soft Spheres system \cite{wesoft-old}. It was found
that Stockes-Einstein relation can be fulfilled there if one takes
the effective particle diameter as Barker perturbation theory one
$d\sim T^{-1/n}$, where $n$ is the softness coefficient. However,
in the case of core-softened systems the situation is more
complicated. These systems have two character length scales and
the applicability of perturbation theory to such systems is
questionable.

From figs.~\ref{fig:fig9} (a) and (b) one can see that the
qualitative behavior of $c_{SE}$ is defined by the viscosity
behavior. For example, the maximum of $c_{SE}$ appears at the
temperatures of minimum of viscosity. At the same time viscosity
of SRSS is monotonous and we observe that $c_{SE}$ is also
monotonous in this case. However, one needs more detailed
investigations to give any conclusive statements on
Stockes-Einstein relation for core-softened systems.

\section{Conclusions}

It is well known from the literature that many core-softened
liquids demonstrate some kind of anomalies. One of the typical
anomalies in the core-softened systems is diffusion anomaly. It is
also widely believed that diffusion is strongly connected to shear
viscosity by Stockes-Einstein relation. In this respect it is
interesting to see if the same systems demonstrate viscosity
anomaly as well. In the present work we investigate this question.
We find that the viscosity anomaly does exist, however, the region
of ($\rho - T$) parameters where viscosity demonstrate anomalous
behavior is different from the diffusion anomaly region. We place
the regions of different anomalies in the same plot to see the
relations between them. Finally we check the Stockes-Einstein
relation for the liquids under investigation.

\bigskip

We thank S. M. Stishov, E. E. Tareyeva and V.V. Brazhkin for
stimulating discussions. Y.F. thanks the Joint Supercomputing
Center of the Russian Academy of Sciences for computational power
and the Russian Scientific Center Kurchatov Institute for
computational facilities. The work was supported in part by the
Russian Foundation for Basic Research (Grants No 13-02-00913, No
11-02-00341-a and No 13-02-00579) the Ministry of Education and
Science of Russian Federation (projects 8370, 8512, Scientific
School No 5365.2012.2 and Young Candidates Grant No 2099.2013.2).



\end{document}